\documentclass[pra,showpacks, showkeys, reprint,amsmath,amssymb,aps]{revtex4-2}
\usepackage{graphicx}
\usepackage{dcolumn}
\usepackage{bm}
\usepackage{epstopdf}
\usepackage{nccmath}
\usepackage{float}
\usepackage{multirow}
\usepackage{appendix}
\usepackage{balance}
\usepackage{silence}
\newcommand{\bra}[1]{\langle#1|} 
\newcommand{\ket}[1]{|#1\rangle} 
\newcommand{\braket}[2]{ \langle #1 | #2 \rangle} 

\begin{document}
	
	\preprint{APS/123-QED}
	
	\title{Modified Six State Cryptographic Protocol with Entangled Ancilla States}
	
	\author{Rashi Jain}
	\email[]{rashijain\_23phdam07@dtu.ac.in}
	\author{Satyabrata Adhikari}
	\email[]{satyabrata@dtu.ac.in}
	\affiliation{Department of Applied Mathematics,\\ Delhi Technological University, Delhi-110042, India}
	%
	\begin{abstract}
		In a realistic situation, it is very difficult to communicate securely between two distant parties without introducing any disturbances. These disturbances might occur either due to external noise or may be due to the interference of an eavesdropper. In this work, we consider and modify the six-state Quantum Key Distribution (QKD) protocol in which Eve can construct the unitary transformation that make all ancilla states entangled at the output, which is not considered in Bruss's work \cite{bruss_1998}. Using the above proposed modification, we would like to study the effect of entangled ancilla states on the mutual information between Alice and Eve. To achieve this task, we calculate the mutual information between Alice and Bob and Alice and Eve, and identify the region where the secret key is generated even in the presence of Eve. We find that, in general, the mutual information of Alice and Eve depends not only on the disturbance D but also on the concurrence of the ancilla states. We show that the entanglement of the ancilla states help in generating the secret key in the region where Bruss's six state QKD protocol failed to do so.  We have further shown that it is possible to derive the disturbance-free mutual information of Alice and Eve, if Eve manipulates her entangled ancilla state in a particular manner. Thus, in this way, we can show that a secret key can be generated between Alice and Bob even if the disturbance is large enough.

	\end{abstract}
	
	\pacs{98.80.-k, 98.80.Es}
	\keywords{Mutual Information, Entanglement, Concurrence, QKD, Disturbance, Secret Key}
	\maketitle{}
	
	\section{Introduction}
	
	One of the most important challenges of the information era is the ability to communicate secretly. QKD \cite{gisin_2002_rev, portmann_2022} is the most advanced application of quantum mechanics \cite{ballentine_2014} and information theory \cite{cover_1999_book, arndt_2001_book}, wherein a secret key is established between two trusted parties using which information can be transmitted secretly. The distribution of the secret key is the most challenging task in the field of quantum cryptography \cite{hklo_1998_book}, as there might be an eavesdropper present between the sender and receiver, assumed to possess every possible operation acceptable by the laws of quantum mechanics \cite{griffith_book}. The security of encrypted messages depends directly on the security of key distribution protocols. QKD uses the most powerful features and resources of quantum mechanics, such as quantum entanglement \cite{horodecki,pirandola_2020_rev} and quantum teleportation \cite{benett95, verstraete_2003,alber_2003_book}. QKD protocols motivated the development of quantum cryptographic protocol variants, including quantum secret sharing \cite{hillery_1999} and quantum-secured blockchain \cite{kiktenko_2018}.\\
	Bennett and Brassard \cite{bb84} were the first to introduce a revolutionary idea of quantum key distribution protocol, which exploits the properties of the quantum mechanics \cite{ashok_2012_book}. They designed a four-state quantum cryptographic protocol for secure key distribution between two distant parties. In this scheme, the sender (Alice) transmits four single qubits in a random manner to the receiver (Bob). The qubit may be prepared either in the computational bases $B_1=\{\ket{0}, \ket{1}\}$ or in the conjugate bases $B_2=\{\ket{\bar{0}},\ket{\bar{1}}\}$, \cite{neilsen_chuang} defined as 
	\begin{equation}
		\begin{split}
			\ket{\bar{0}}&=\frac{1}{\sqrt{2}}(\ket{0}+\ket{1})\\
			\ket{\bar{1}}&=\frac{1}{\sqrt{2}}(\ket{0}-\ket{1})	
		\end{split}
		\label{plus_minus basis}
	\end{equation}
	Bob performs measurements on one of the two bases mentioned above at random. Then, in the next step, the bases used by both, Alice and Bob are disclosed in the public channel where they compare the bit values for a randomly chosen sample of few bits from the string prepared by both sender and receiver. A secret key is said to be established using the cases in which Alice's and Bob's bases coincide. It is then important for Alice and Bob to know whether anyone in the middle may steal the information about the key. Alice and Bob can reveal the existence of the third party, i.e., the eavesdropper in the middle, by calculating the quantum bit error rate (QBER). QBER may be calculated as the ratio of the error rate to the key rate. Therefore, if the QBER is too high, Alice and Bob decide to abort the protocol. Otherwise, using either one-way \cite{stucki_2005,kraus_2005,renner_2005} or two-way \cite{gottesman_2003, beaudry_2013} classical communication, they apply a classical post-processing protocol to distill a secret key.\\  
	The security of the standard BB84 scheme has been provided in \cite{mayars_2001}, which is closer to a realistic experimental situation. Later, simple proof of security of BB84 was presented in \cite{hklo_1999,hklo_2001,shor_perskill_2000}. A general framework of optimal eavesdropping on BB84 protocol was studied in \cite{fuchs_1997} and they derived an upper bound on mutual information and described a specific type of interaction and the corresponding measurement that achieves the bound. Extending their work, in 2017, the uniqueness of optimal interaction up to rotation has been established in \cite{atanu_2017}.\\
	In 1998, D. Bruß \cite{bruss_1998} introduced a generalization of the BB84 scheme in the sense that the author uses three bases i.e. six states in her QKD protocol instead of four states used in BB84 protocol. The third set of bases may be denoted as $B_{3}$ in which the states can be defined as 
	\begin{equation}
		\begin{split}
			\ket{\bar{\bar{0}}}&=\frac{1}{\sqrt{2}}(\ket{0}+i \ket{1})\\
			\ket{\bar{\bar{1}}}&=\frac{1}{\sqrt{2}}(\ket{0}-i \ket{1})
		\end{split}
		\label{iota basis}
	\end{equation}
	In the Bruß scheme \cite{bruss_1998}, Alice transmits one of these six states to Bob, with equal probability, and in this scheme, it is assumed that all six possible states occurred at Bob's site with the same disturbance. Using the optimal mutual information between Alice and Eve, it was shown that this scheme \cite{bruss_1998} is more secure against eavesdropping in comparison to BB84 scheme \cite{bb84}. In 2008, an optimal six-state eavesdropping QKD scheme \cite{shadman_2009} was proposed wherein Alice deliberately adds equal noise to the signal states before transmitting it to Bob, and they found that the six-state protocol with mixed states is more robust than the six-state protocol with pure states. The proof of security of the six-state QKD protocol has been proposed by \cite{inamori_2000,scarani_2009}. \\ 
	Moreover, we can also find that the security increases not only by increasing the number of non-orthogonal bases but also by increasing the level of the system.  This can be easily verified from the study, where Bechmann and Peres \cite{bechmann_peres} suggested the use of three-level systems rather than two-level systems for establishing a secure quantum key. Later, in 2002, Bruß and Macchiavello \cite{bruss_2002} introduced the optimal eavesdropping scheme with three-dimensional systems, and showed that it would be more secure against symmetric attacks than two-dimensional systems. One can also note that, as the number of bases increases, the security also increases, but the efficiency of key formation decreases \cite{bruss_2002}. Further, in 1999, Gisin and Wolf \cite{gisin_wolf_1999} studied a protocol in which as long as Alice and Bob share some entanglement, they can use either quantum or classical key-agreement protocols to extract a secret key.\\
	Despite of all these studies related to QKD protocols, we observe that there is a gap in which we can think of a situation, where Eve can execute her unitary operation in such a way that she may  extract the information as much as possible successfully from the intercepted qubit by allowing the small disturbance and thus omitting her presence in the protocol. Hence, this gap motivate us to study about Eve, a person in the middle, who believed to be stronger in the sense that she can use quantum resource efficiently.\\ 
	The aim of this paper is two-fold. Firstly, we observe that the ancilla states included in the unitary transformation introduced by Bruss \cite{bruss_1998} are not all entangled. The question here arises that if all the ancilla states are entangled, then what effect will it have on the mutual information between Alice and Eve. Thus, our aim is to show that if Eve construct the unitary transformation in such a way that it may transform her input ancilla state to the entangled ancilla states at the output. With the help of this construction, our purpose is to show that the mutual information of Alice and Eve may depend not only on the disturbance but also on the concurrence of the entangled ancilla states. Secondly, our aim is to show that the generation of the secret key also depends on the entanglement of ancilla states.\\  
	The rest of the paper is organized as follows: In Section II, we will discuss in detail the unitary transformation constructed by Eve that may generate the entangled ancilla states at the output of the transformation. In section III, we derive the mutual information of Alice and Eve and show that it depends on the concurrence of the entangled ancilla states. Moreover, we also derive the mutual information of Alice and Bob, which solely depends on the disturbance $D$. In section IV, we discuss the generation of the secret key in the following two cases: (i) when the mutual information of Alice and Eve depends on the disturbance $D$ caused by Eve's interference and (ii) when it is independent of the same. Finally, we conclude in section V.   
	
	\section{Construction of a Unitary transformation }
	In a six state QKD protocol, the sender (Alice) has to send her qubit prepared in any one of the basis chosen from $\{B_1,B_2,B_3\}$ to the receiver (Bob). But before reaching to Bob, the eavesdropper (Eve) intercepts the qubit with the following motivations: (i) Eve would like to extract as much information as possible from the qubit sent by Alice so that she would know about the secret key shared between Alice and Bob (ii) While extracting the information from the intercepting qubit, she also has to keep in mind that the disturbance will be the same irrespective of the basis $\{B_1,B_2,B_3\}$ chosen for the preparation of Alice's qubit. This ensures the fact that Bob may not discover the presence of an Eavesdropper.\\
	Thus, to achieve her motivations, Eve has to apply a suitable unitary transformation on the intercepted qubit. Therefore, it is crucial to consider all the above points while constructing the unitary transformation. The unitary transformation $U$ constructed by Eve may be defined as
	\begin{equation}
		\begin{split}
			U\ket{0}\ket{X}=\sqrt{F}\ket{0}\ket{\phi_0}+\sqrt{D}\ket{1}\ket{\psi_0}\\
			U\ket{1}\ket{X}=\sqrt{F}\ket{1}\ket{\phi_1}+\sqrt{D}\ket{0}\ket{\psi_1}
		\end{split}		
		\label{transformation}
	\end{equation}
	where $|X\rangle$ denote the initial normalized ancilla state before interaction with the input qubit while $\ket{\phi_i}$ and $\ket{\psi_i}$ $(i=0,1)$ refers to  normalized ancilla states after interaction. $D$ denote the disturbance at the output when the ancillary state of Eve interacted with the input qubit. The fidelity between the input and output qubit is described by $F$ and it is related with $D$ as given below
	\begin{equation}
		F+D=1
		\label{fisdist}
	\end{equation}
	\subsection{Constraints to be satisfied by the ancilla states at the output}
	The constraints that can be satisfied by the ancilla states at the output might come from two assumptions: (a) the unitarity of the transformation and (b) the transformation should keep the disturbance $D$ same for all input qubits that come from any basis under consideration.\\  
	(a) Unitarity of the transformation gives
	\begin{align}
		\braket{\phi_0}{\psi_1}+\braket{\psi_0}{\phi_1}&=0
		\label{unitaryU}
	\end{align}
	(b) If the transformation is treating all the input states in the same way i.e. keeping the disturbance same for all input qubits generated by sender from any basis, then the following conditions hold
	\begin{align}
		\braket{\psi_0}{\psi_1}&=0 \label{bd}\\
		\text{Re}\braket{\phi_1}{\phi_0}&=\frac{1-2D}{1-D}\label{reac}\\
		\braket{\phi_0}{\psi_0}+\overline{\braket{\phi_1}{\psi_1}}&=0 \label{constraint}
	\end{align}

	\subsection{Structure of the two-qubit ancilla states}
	We are now in a position to determine the structure of the two-qubit ancilla states generated at the output when the initial ancilla states of Eve interact with the qubit sent by the sender, Alice. The specific form of the ancilla states may be obtained by using the constraints given in (\ref{bd}-\ref{constraint}). 
	\subsubsection{Specific form of $\ket{\psi_0}$ and $\ket{\psi_1}$ }
	To start with, let us first choose the ancilla state $\ket{\psi_0}$ and $\ket{\psi_1}$ in such a way that (\ref{bd}) holds. To fulfill the requirement, we may choose
	\begin{align}
		\ket{\psi_0}&=\frac{\ket{00}-\tau_1\ket{11}}{\sqrt{1+\tau_1^2}}
		\label{ket B}
	\end{align}
	and 
	\begin{align}
		\ket{\psi_1}&=\frac{\tau_1\ket{00}+\ket{11}}{\sqrt{1+\tau_1^2}}
		\label{ket D}
	\end{align}
	where $0\leq \tau_1 \leq 1$.\\
	Defining in this way (\ref{ket B}-\ref{ket D}), we can say that the ancilla states $\ket{\psi_0}$ and $\ket{\psi_1}$ represents non-maximally entangled states. They are product states when $\tau_1=0$, while maximally entangled states when $\tau_1=1$. 
	To quantify the amount of entanglement in the states $\ket{\psi_0}$ and $\ket{\psi_1}$, we can make use of concurrence \cite{wootters_2001}. The concurrence $C_{\psi_0}$ and $C_{\psi_1}$ of the state $\ket{\psi_0}$ and $\ket{\psi_1}$ respectively may be given as
	\begin{align}
		C_{\psi_0}&=C_{\psi_1}=\frac{2 \tau_1}{1+\tau_1^2}
		\label{cb}
	\end{align}
	From (\ref{cb}), we can express $\tau_1^2$ in terms of  $C_{\psi_0}$ as
	\begin{align}
		\tau_1^2= \frac{2-C_{\psi_0}^2- 2\sqrt{1-C_{\psi_0}^2}}{C_{\psi_0}^2}
		\label{tau1sq}
	\end{align}
	The proof of (\ref{tau1sq}) is given in Appendix-A.\\
	We should note here that when $\tau_1=0$, the transformation (\ref{transformation}) reduces to the transformation prescribed by Bruss \cite{bruss_1998}. Here, we may also stress that when $0 < \tau_1 \leq 1$, the transformation may shed new light on the distribution of the secret key between the sender and the receiver. We will show that the formation of a secret key is possible, although the modified transformation may increase the efficiency of Eve with respect to the stealing of information.
	\subsubsection{Specific form of $\ket{\phi_0}$ and $\ket{\phi_1}$ }
	Let us now move on to the ancilla states $\ket{\phi_0}$ and $\ket{\phi_1}$, which can be expressed initially in the general form of two-qubit pure states in the computational basis $\{\ket{00},\ket{01},\ket{10},\ket{11}\}$ as 
	\begin{align}
		\ket{\phi_0}&=\alpha_{\phi_0} \ket{00}+\beta_{\phi_0} \ket{01}+\gamma_{\phi_0} \ket{10}+\delta_{\phi_0} \ket{11} \label{ketAorg}\\
		\text{with}& ~|\alpha_{\phi_0}|^2+|\beta_{\phi_0}|^2+|\gamma_{\phi_0}|^2+|\delta_{\phi_0}|^2=1\notag 
	\end{align}
	and
	\begin{align}
		\ket{\phi_1}&=\alpha_{\phi_1} \ket{00}+\beta_{\phi_1} \ket{01}+\gamma_{\phi_1} \ket{10}+\delta_{\phi_1} \ket{11}\label{ketCorg}\\
		\text{with} & ~|\alpha_{\phi_1}|^2+|\beta_{\phi_1}|^2+|\gamma_{\phi_1}|^2+|\delta_{\phi_1}|^2=1 \notag		
	\end{align}
	Now, our task is to determine the parameters $\alpha_{\phi_i}, \beta_{\phi_i}, \gamma_{\phi_i}$ and $\delta_{\phi_i}$ $(i=0,1)$, for which the constraints (\ref{unitaryU}-\ref{constraint}) are satisfied. Simplifying the constraints (\ref{unitaryU}-\ref{constraint}) for the states $\ket{\phi_0}$ and $\ket{\phi_1}$, we get (Appendix B) 
	\begin{align}
		\alpha_{\phi_0}=\alpha_{\phi_1}=0
		\label{alpha}
	\end{align}	
	
	\begin{align}
		\delta_{\phi_0}=\delta_{\phi_1}=0
		\label{delta}
	\end{align}	
	Therefore, the ancilla states $\ket{\phi_0}$ and $\ket{\phi_1}$ given in (\ref{ketAorg}) and (\ref{ketCorg}) reduces to 
	\begin{align}
		\ket{\chi_0}&= \beta_{\phi_0} \ket{01}+\gamma_{\phi_0}\ket{10}, ~ |\beta_{\phi_0}|^2+|\gamma_{\phi_0}|^2=1
		\label{ket A}
	\end{align}
	\begin{align}
		\ket{\chi_1}&= \beta_{\phi_1} \ket{01}+\gamma_{\phi_1}\ket{10}, ~ |\beta_{\phi_1}|^2+|\gamma_{\phi_1}|^2=1 
		\label{ket C}
	\end{align}	
	It may be easily seen that the ancilla states $\ket{\chi_0}$ and $\ket{\chi_1}$ are entangled unless the parameters $|\beta_{\phi_0}|$ or $|\gamma_{\phi_0}|$ and  $|\beta_{\phi_1}|$ or $|\gamma_{\phi_1}|$ vanishes. If they are entangled, then the amount of entanglement can be calculated by concurrence, and it is given by
	\begin{align}
		C_{\chi_0}&=2|\text{Re}(\beta_{\phi_0}\gamma^*_{\phi_0})|\notag \\
		&=2|b_1g_1+b_2g_2|
		\label{concphi}
	\end{align}
	where $\beta_{\phi_0}=b_1+i b_2$ and $\gamma_{\phi_0}=g_1+i g_2$.\\	
	If we now choose the parameters $b_1,b_2,g_1$ and $g_2$ in such a way that $b_1g_2=b_2g_1$ then the equation (\ref{concphi}) reduces to
	\begin{align}
		C_{\chi_0}^{2}&=4\Big(|\beta_{\phi_0}|^2(1-|\beta_{\phi_0}|^2)\Big)
		\label{C_A2}
	\end{align}	
	Solving (\ref{C_A2}), we can obtain the value of $|\beta_{\phi_0}|^2$ in terms of the concurrence $C_{\chi_0}$ as
	\begin{align}
		|\beta_{\phi_0}|^2 &=\frac{1\pm \sqrt{1-C_{\chi_0}^{2}}}{2}, 0\leq C_{\chi_0} \leq 1
		\label{beta}
	\end{align}
	One of the two values of $|\beta_{\phi_0}|^2$ can be considered from (\ref{beta}) and then the ranges of $|\beta_{\phi_0}|^2$ can be found out, which are discussed below:\\
	(i) If 
	\begin{equation}
		|\beta_{\phi_0}|^2 =\frac{1-\sqrt{1-C_{\chi_0}^{2}}}{2} 
		\label{value1}
	\end{equation}
	then we can find that $|\beta_{\phi_0}|^2$ is lying in $[0,\frac{1}{2}]$. We may also observe that $|\beta_{\phi_0}|^2$ is an increasing function of $C_{\chi_0}$.\\
	(ii) If 
	\begin{equation}
		|\beta_{\phi_0}|^2 =\frac{1+\sqrt{1-C_{\chi_0}^{2}}}{2}
		\label{value2}
	\end{equation}
	then $|\beta_{\phi_0}|^2 \in [\frac{1}{2},1]$. In this case, we find that $|\beta_{\phi_0}|^2$ is a decreasing function of $C_{\chi_0}$.\\
	In a similar fashion, the concurrence of the ancilla state $\ket{\chi_1}$ can be denoted by $C_{\chi_1}$, which is given by
	\begin{align}
		C^{2}_{\chi_{1}}&=4\Big(|\beta_{\phi_1}|^2(1-|\beta_{\phi_1}|^2)\Big)
		\label{CA2}
	\end{align}
	Therefore, $|\beta_{\phi_1}|^2$ can be expressed in terms of the concurrence $C_{\chi_1}$ as
	\begin{align}
		|\beta_{\phi_1}|^2 &=\frac{1\pm \sqrt{1-C_{\chi_1}^{2}}}{2}, 0\leq C_{\chi_1} \leq 1
		\label{ba2 and ca}
	\end{align}	
	Following the similar discussion above, we can have  
	\begin{equation}
		\begin{split}
			(i)~~ \text{If}~~ |\beta_{\phi_1}|^2 &=\frac{1- \sqrt{1-C^{2}_{\chi_{1}}}}{2}~~ \text{then}~~ |\beta_{\phi_1}|^2 \in \bigg[0,\frac{1}{2}\bigg]\\
			(ii)~~\text{If}~~|\beta_{\phi_1}|^2 &=\frac{1+\sqrt{1-C^{2}_{\chi_{1}}}}{2} ~~ \text{then} ~~|\beta_{\phi_1}|^2 \in \bigg[\frac{1}{2},1\bigg]
			\label{conc123}
		\end{split}
	\end{equation}
	
	\section{Mutual Information}
	Mutual information can be defined as the relative entropy between the joint distribution and the product of the marginal distributions. If $p(x,y)$ denote the joint probability mass function of the random variables $X$ and $Y$ and if $p(x)$ and $p(y)$ denote the marginal probability mass functions then the mutual information may be defined mathematically as \cite{cover_1999_book,mceliece_2002_book}
	\begin{align}
		I^{XY}= \sum_{x}\sum_{y}p(x,y)\log_{2}\frac{p(x,y)}{p(x)p(y)}
		\label{mutualinf}
	\end{align}	
	Mutual information $I^{XY}$ may also be interpreted as how much information we can extract about the random variable $X$ by performing a measurement on other random variable $Y$.\\ 
	Therefore, our interest in this section would be to know how well Eve will use her entangled ancilla states to extract information from the intercepted qubit, which was sent by the sender, Alice. \\ 
	\subsection{Mutual information of Alice and Eve}
	Let Alice prepares a two-qubit maximally entangled state of the form 
	\begin{align}
		\ket{\psi^{(B_1)}}=\frac{1}{\sqrt{2}}(\ket{00}+\ket{11})
		\label{maxent_b1}
	\end{align}	
	Here, we should note that it is not necessary that Alice prepare the qubit in only computational basis $B_{1}$= $\{ \ket{0},\ket{1}\}$ but she may use the basis $B_{2}$= $\{ \frac{\ket{0}+\ket{1}}{\sqrt{2}},\frac{\ket{0}-\ket{1}}{\sqrt{2}}\}$ or $B_{3}$= $\{ \frac{\ket{0}+i\ket{1}}{\sqrt{2}},\frac{\ket{0}-i\ket{1}}{\sqrt{2}}\}$ also.
	Using the bases $B_2$ and $B_3$, the two-qubit maximally entangled state can be re-expressed as 
	\begin{align}
		\ket{\psi^{(B_2)}}=&\frac{1}{\sqrt{2}}\bigg[(\ket{0}+\ket{1})(\ket{0}+\ket{1})+(\ket{0}-\ket{1})(\ket{0}-\ket{1})\bigg]
		\label{maxent_b2}
	\end{align}	
	\begin{align}
		\ket{\psi^{(B_3)}}=&\frac{1}{\sqrt{2}}\bigg[(\ket{0}+i\ket{1})(\ket{0}+i\ket{1})+(\ket{0}-i\ket{1})(\ket{0}-i\ket{1})\bigg]
		\label{maxent_b3}
	\end{align}	
	
	She then transmits one of the qubits to Bob, but in the midway, Eve intercepts the qubit, and the intercepted qubit undergoes the transformation (\ref{transformation}). Eve does not know the fact that in which basis, Alice has prepared her qubits. But Eve's transformation will act on the qubit in the same way irrespective of the basis chosen by the sender, Alice, to prepare the state. This means that the disturbance created by Eve's transformation will be same irrespective of the chosen basis.\\
	When the intercepted qubit interacts with Eve's ancilla state, the four-qubit state at Eve's site is given by
	\begin{align}
		\ket {\chi}_{AEE_1E_2}&=\frac{1}{\sqrt{2}}\bigg(\sqrt{F}\ket{00}_{AE}\ket{\phi_0}_{E_1E_2}+\sqrt{D}\ket{01}_{AE}\ket{\psi_0}_{E_1E_2}\notag \\
		&+\sqrt{F}\ket{11}_{AE}\ket{\phi_1}_{E_1E_2} +\sqrt{D}\ket{10}_{AE}\ket{\psi_1}_{E_1E_2}\bigg)
	\end{align}
	where the first qubit possessed by Alice, the second qubit is in Eve's possession and the remaining two-qubits are from Eve's ancilla state.\\
	Now, tracing out the two ancilla qubits $E_{1}$ and $E_{2}$, the reduced state shared between Alice and Eve is given by 
	\begin{small}
		\begin{align}
			\begin{split}
				\rho_{AE} &=  \frac{1}{2}[\left(F|\beta_{\phi_0}|^2+\eta\right)\ket{00}\bra{00} + \sqrt{F\eta}\left(\beta_{\phi_0}+\gamma^*_{\phi_1}\right)\ket{00}\bra{11}\\
				+& \left(F|\gamma_{\phi_0}|^2+\eta\tau_1^2\right)\ket{01}\bra{01} + \sqrt{F\eta}\tau_1^2\left(\gamma_{\phi_0}-\beta^*_{\phi_1}\right)\ket{01}\bra{10}\\
				+& \sqrt{F\eta}\tau_1^2\left(\beta_{\phi_1}+\gamma^*_{\phi_1}\right)\ket{10}\bra{01}+ \left(F|\beta_{\phi_1}|^2+\eta\tau_1^2\right)\ket{10}\bra{10}\\
				+& \sqrt{F\eta}\left(\gamma_{\phi_1}+\beta^*_{\phi_0}\right)\ket{11}\bra{00} + \left(F|\gamma_{\phi_1}|^2+\eta\right)\ket{11}\bra{11}]
			\end{split}
		\end{align}
	\end{small}
	where $\eta=\frac{D}{1+\tau_1^2}$.\\
	The mutual information of Alice and Eve is given by
	
	\begin{align}
		I^{AE}&(|\beta_{\phi_0}|^2,|\beta_{\phi_1}|^2,\tau_1,D)= F(|\beta_{\phi_0}|^2+|\beta_{\phi_1}|^2)+D \notag-\\
		&\left(F(|\beta_{\phi_0}|^2+|\beta_{\phi_1}|^2)+D\right) \log_2 \left(F(|\beta_{\phi_0}|^2+|\beta_{\phi_1}|^2)+D\right)\notag\\
		&+\epsilon\big[F|\beta_{\phi_0}|^2+\eta, F|\beta_{\phi_1}|^2+\eta \tau_1^2\big]
		\label{iae}
	\end{align}
	where the parameters $|\beta_{\phi_0}|^2$, $|\beta_{\phi_1}|^2$ are defined in (\ref{value1}), (\ref{value2}) and (\ref{conc123}) and the function $\epsilon[a,b]$ may be defined as
	\begin{equation}
		\epsilon[a,b]=a \log a+ b \log b -(a+b) \log (a+b)
		\label{tau_def}
	\end{equation}
	
	Therefore, the mutual information between Alice and Eve may be re-expressed in terms of the concurrences $C_{\chi_0}$ and $C_{\chi_1}$ as
	\begin{widetext}
		\begin{small}
			\begin{align}
				(i)~~ I^{AE}_1(C_{\chi_0},C_{\chi_1},\tau_1,D)&=F\left(\frac{1-\sqrt{1-C_{\chi_0}^2}}{2} +\frac{1-\sqrt{1-C_{\chi_1}^2}}{2}\right)+D+\epsilon\bigg[F\left(\frac{1-\sqrt{1-C_{\chi_0}^2}}{2}\right) +\eta, F\left(\frac{1-\sqrt{1-C_{\chi_1}^2}}{2}\right)+\eta\tau_1^2\bigg],\notag\\ 
				&|\beta_{\phi_0}|^2\in \bigg[0,\frac{1}{2}\bigg], |\beta_{\phi_1}|^2 \in \bigg[0,\frac{1}{2}\bigg]\\
				(ii)~~ I^{AE}_2(C_{\chi_0},C_{\chi_1},\tau_1,D)&=F\left(\frac{1-\sqrt{1-C_{\chi_0}^2}}{2} +\frac{1+\sqrt{1-C_{\chi_1}^2}}{2}\right)+D+\epsilon\bigg[F\left(\frac{1-\sqrt{1-C_{\chi_0}^2}}{2}\right) +\eta, F\left(\frac{1+\sqrt{1-C_{\chi_1}^2}}{2}\right)+\eta\tau_1^2\bigg],\notag\\
				&|\beta_{\phi_0}|^2\in \bigg[0,\frac{1}{2}\bigg], |\beta_{\phi_1}|^2 \in \bigg[\frac{1}{2},1\bigg]\\
				(iii)~~ I^{AE}_3(C_{\chi_0},C_{\chi_1},\tau_1,D)&=F\left(\frac{1+\sqrt{1-C_{\chi_0}^2}}{2} +\frac{1-\sqrt{1-C_{\chi_1}^2}}{2}\right)+D+\epsilon\bigg[F\left(\frac{1+\sqrt{1-C_{\chi_0}^2}}{2}\right) +\eta, F\left(\frac{1-\sqrt{1-C_{\chi_1}^2}}{2}\right)+\eta\tau_1^2\bigg],\notag\\
				&|\beta_{\phi_0}|^2\in \bigg[\frac{1}{2},1\bigg], |\beta_{\phi_1}|^2 \in \bigg[0,\frac{1}{2}\bigg]\\
				(iv)~~ I^{AE}_4(C_{\chi_0},C_{\chi_1},\tau_1,D)&=F\left(\frac{1+\sqrt{1-C_{\chi_0}^2}}{2} +\frac{1+\sqrt{1+C_{\chi_1}^2}}{2}\right)+D+\epsilon\bigg[F\left(\frac{1+\sqrt{1-C_{\chi_0}^2}}{2}\right) +\eta, F\left(\frac{1+\sqrt{1-C_{\chi_1}^2}}{2}\right)+\eta\tau_1^2\bigg]\notag\\
				&|\beta_{\phi_0}|^2\in \bigg[\frac{1}{2},1\bigg], |\beta_{\phi_1}|^2 \in \bigg[\frac{1}{2},1\bigg]
			\end{align}
		\end{small}
	\end{widetext}

\subsection{Mutual information of Alice and Bob}
Eve extract the information from the intercepted qubit by using her ancilla states in such a way that it keeps the disturbance same for all the intercepted qubit irrespective of the bases in which Alice has prepared the qubit. After eavesdropping, Eve sent the intercepted qubit to Bob so that Alice and Bob does not reveal her presence in the intermediate place.
After receiving the qubit, the shared state between Alice and Bob is given by
\begin{align}
	\rho_{AB}&=	\begin{pmatrix}
		\frac{1-D}{2} & 0 &0  & \frac{1-2D}{4}  \\
		0	& \frac{D}{2} &  0&  0\\
		0	& 0 & \frac{D}{2} &0  \\
		\frac{1-2D}{4}	&0  &0  & \frac{1-D}{2}
	\end{pmatrix}
	\label{rhoAB}
\end{align}
Therefore, the mutual information of Alice and Bob is given by
\begin{equation}
I^{AB}(D)= 1+D \log_2 D +(1-D) \log_2 (1-D)
\end{equation}

\section{Generation of Secret Key in six state modified QKD protocol}
In this section, we will provide a detailed analysis of when the secret key will be generated in the quantum key distribution protocol in spite of the presence of an eavesdropper with entangled ancilla states. It is known that the two parties can share a secret key if the mutual information of $A$ and $B$ is strictly greater than the mutual information of $A$ and $E$ \cite{csiszar_korner}. To analyze it in detail, we divide the whole fact into two parts: In the first part, we will investigate the generation of the secret key when the mutual information of Alice and Eve depends on the disturbance $D$, and in the second part, we probe when mutual information of Alice and Eve is independent of $D$.

\subsection{Relationship between $I^{AE}(C_{\psi_0},C_{\chi_0},D)$ and the concurrences $C_{\psi_0}$ and $C_{\chi_0}$ of the ancilla states}
Eve can enhance the amount of extracted information from the intercepted qubit if she chooses the parameter $|\beta_{\phi_0}|^2$ and $|\beta_{\phi_1}|^2$ in (\ref{iae}) in the following way:
\begin{align}
|\beta_{\phi_0}|^2+|\beta_{\phi_1}|^2=1
\label{cons1}
\end{align}
Therefore, using (\ref{fisdist}) and (\ref{cons1}), the mutual information $I^{AE}(|\beta_{\phi_0}|^2,|\beta_{\phi_1}|^2,\tau_1,D)$ given in (\ref{iae}) reduces to
\begin{align}
I^{AE}(|\beta_{\phi_0}|^2,\tau_1,D)&= 1+\left[(1-D)|\beta_{\phi_0}|^2+\eta\right]\times \notag\\
&\log_2 \left[(1-D)|\beta_{\phi_0}|^2+\eta\right] + \notag \\
&\left[(1-D)(1-|\beta_{\phi_0}|^2)+\eta\tau_1^2\right]\times \notag\\
&\log_2 \left[(1-D)(1-|\beta_{\phi_0}|^2)+\eta\tau_1^2\right]
\label{iaered}
\end{align}
The analysis of the generation of secret key can be studied in the following two cases:\\

\textbf{Case-I:} When $|\beta_{\phi_0}|^2\in [0,\frac{1}{2}]$.\\
In this case, the relationship between $|\beta_{\phi_0}|^2$ and the concurrence $C_{\chi_0}^{2}$ of the ancilla state $\ket{\chi_0}$ is given by using (\ref{tau1sq}) and (\ref{value1}), the expression of $I^{AE}(|\beta_{\phi_0}|^2,\tau_{1}^{2},D)$ given in (\ref{iaered}) reduces to
\begin{widetext}
\begin{align}
	I^{AE}(C_{\psi_0},C_{\chi_0},D)&=1+ \frac{1}{2} \left((1-\sqrt{1-C_{\chi_0}^{2}}) +\frac{D\left(C_{\psi_0}^2-(1-\sqrt{1-C_{\chi_0}^{2}})(1-\sqrt{1-C_{\psi_0}^{2}})\right)}{1-\sqrt{1-C_{\psi_0}^{2}}}\right)\notag\\
	&\log_2 \left[\frac{1}{2} \left((1-\sqrt{1-C_{\chi_0}^{2}}) +\frac{D\left(C_{\psi_0}^2-(1-\sqrt{1-C_{\chi_0}^{2}})(1-\sqrt{1-C_{\psi_0}^{2}})\right)}{1-\sqrt{1-C_{\psi_0}^{2}}}\right) \right]+\notag\\
	&\frac{1}{2}\left((1+\sqrt{1-C_{\chi_0}^{2}}) +\frac{D\left(2-C_{\psi_0}^2-2\sqrt{1-C_{\psi_0}^2}-\left(1+\sqrt{1-C_{\chi_0}^{2}}\right)\left(1-\sqrt{1-C_{\psi_0}^{2}}\right)\right)}{1-\sqrt{1-C_{\psi_0}^{2}}} \right)\notag\\
	&\log_2 \left[\frac{1}{2}\left((1+\sqrt{1-C_{\chi_0}^{2}}) +\frac{D\left(2-C_{\psi_0}^2-2\sqrt{1-C_{\psi_0}^2}-\left(1+\sqrt{1-C_{\chi_0}^{2}}\right)\left(1-\sqrt{1-C_{\psi_0}^{2}}\right)\right)}{1-\sqrt{1-C_{\psi_0}^{2}}} \right)\right] 
	\label{Iaed}
\end{align}
\end{widetext}
It may be observed that if we vary the parameters $C_{\chi_0}$ and $C_{\psi_0}$ in between zero and unity and if $D$ is lying between zero and a value which is very close to $0.5$, then we get  
\begin{equation}
I^{AB}(D)-I^{AE}(C_{\psi_0},C_{\chi_0},D)>0,~~
\label{ineq1}
\end{equation}
Thus, equation (\ref{ineq1}) represents the fact that, in this case, a secret key can be generated. But, there exists a critical value of $D$, which is lying in $[0.4999999.0.5)$ and for the concurrence $C_{\psi_0}\in (0,0.009]$, for which a secret key cannot be generated. This can be verified if we choose the values of the concurrences $C_{\chi_0}$ and $C_{\psi_0}$ in such a way that $C_{\chi_0}=C_{\psi_0}$. In this particular case,  $I^{AE}(C_{\psi_0},C_{\chi_0},D)$ given in (\ref{Iaed}) can be re-expressed as
\begin{widetext}
\begin{align}
	I^{AE}(C_{\psi_0},D)=&1+\frac{1}{2}\left(1-(1-2D)\sqrt{1-C_{\psi_0}^2}\right) \log_2 \left [ \frac{1}{2}\left(1-(1-2D)\sqrt{1-C_{\psi_0}^2}\right))\right ]+\notag \\
	&\frac{1}{2}\left(1+(1-2D)\sqrt{1-C_{\psi_0}^2}\right) \log_2 \left [ \frac{1}{2}\left(1+(1-2D)\sqrt{1-C_{\psi_0}^2}\right)\right ]  \label{exp1}
\end{align}
\end{widetext}
Now, in the above expression (\ref{exp1}), if we choose the values of the parameter $C_{\psi_0}$ from $0.1$ to $0.9$ with a step length of $0.1$ and varying $D$ in the interval (0,0.5) then the plot of $I^{AB}(D)-I^{AE}(C_{\psi_0},D)$ versus $D$ may be shown in the Fig.\ref{dependent_on_D}.
\begin{figure}[H]
	\centering
	\includegraphics[width=1\linewidth]{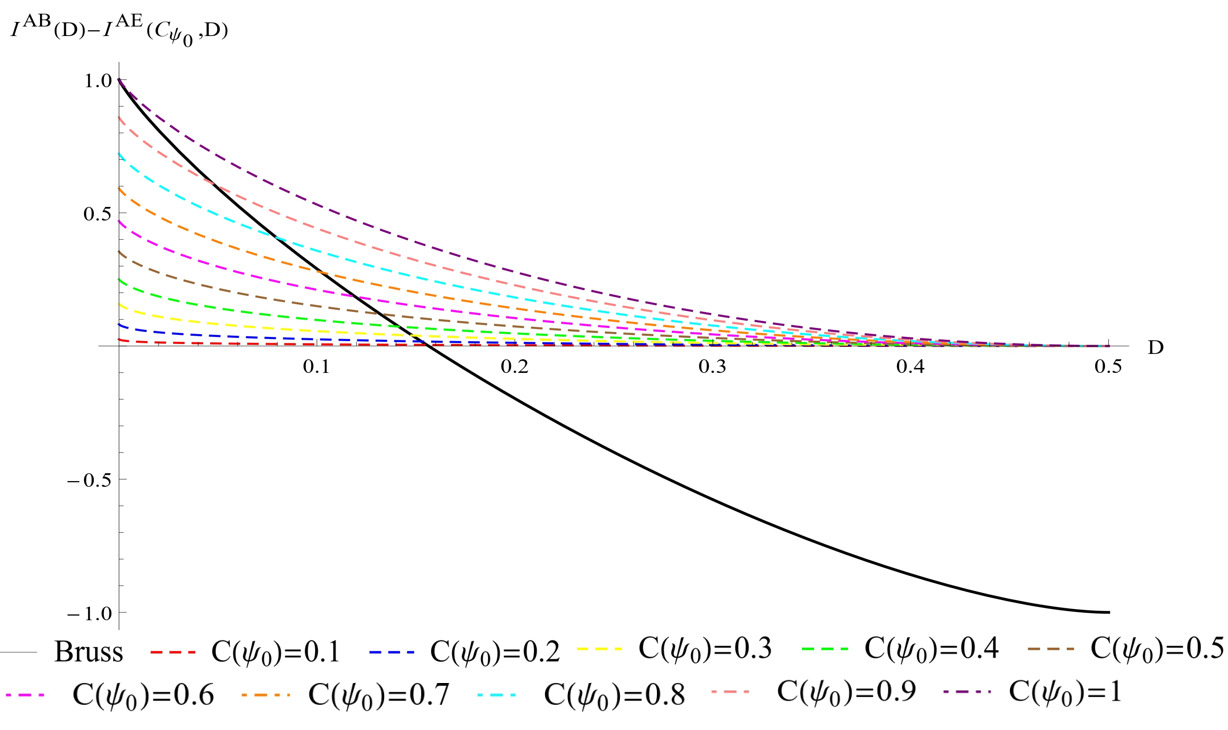}
	\caption{Secret key generation for the six-state modified QKD protocol with each ancilla state as entangled states compared to the six-state protocol adopted by \cite{bruss_1998}. The $D$-axis represents the amount of disturbance $D\in (0,0.5)$ in the system as a result of Eve's interference and the $I^{AB}(D)-I^{AE}(C_{\psi_0},D)$ axis represents the possibility for generation of the secret key between Alice and Bob in Eve's presence.}
	\label{fig:dependent-d-graph}
\end{figure}
The following insights can be drawn from the graph (FIG.\ref{dependent_on_D}) obtained above:\\
(i) As the concurrence $C_{\psi_0}$ increases from 0 to 1, the value of $I^{AB}(D)-I^{AE}(C_{\psi_0},D)$ increases as the disturbance $D$ increases from 0 to 0.5.\\
(ii) After a certain value of disturbance $D$, the key is not generated for the model proposed by \cite{bruss_1998}. However, it can be easily seen that almost all the curves in our case lie in the upper half of the $D$-axis, implying that the secret key can be generated as $I^{AB}(D)-I^{AE}(C_{\psi_0},D)>0$ for all values of $C_{\psi_0}\in (0.009, 1]$ and for the values of $D$ lying in $(0,0.4999999)$. The secret key cannot be generated when $D\in [0.4999999,0.5)$ and $C_{\psi_0}\in (0, 0.009]$. \\

\textbf{Case-II:} When $|\beta_{\phi_0}|^2\in [\frac{1}{2},1]$.\\
Using (\ref{tau1sq}) and (\ref{value2}), we find that the expression given in (\ref{iaered}) will be independent of $D$, and it will be studied in detail in the next subsection.\\ 
\subsection{Relationship between the concurrence $C_{\psi_0}$ of the ancilla states and  $D$ independent mutual information} 	
Here, we discuss the case when there is no effect of $D$ on the mutual information of Alice and Eve. That is, mutual information remains the same for whatever disturbance induced by Eve on the intercepted qubit while extracting information from it. In this scenario, mutual information depends only on the entanglement of the ancilla states.\\
Recalling (\ref{iaered}), let's re-express it in the following way
\begin{align}
I^{AE}(|\beta_{\phi_0}|^2,\tau_{1}^{2},D)&=1+\left[(1-D)|\beta_{\phi_0}|^2+\eta\right]\times \notag \\
&\log_2 \left[(1-D)|\beta_{\phi_0}|^2+\eta\right] \notag\\
&+\left[(1-D)|\beta_{\phi_1}|^2+\eta\tau_1^2\right]\times \notag \\
&\log_2 \left[(1-D)|\beta_{\phi_1}|^2+\eta\tau_1^2\right]
\label{Iae_final}
\end{align}
If we now choose the parameters $|\beta_{\phi_0}|^2$ and $|\beta_{\phi_1}|^2$ in such a way that
\begin{equation}
|\beta_{\phi_0}|^2=\frac{1}{1+\tau_1^2}, |\beta_{\phi_1}|^2=\frac{\tau_1^2}{1+\tau_1^2}
\label{cons2}
\end{equation}
then (\ref{cons1}) holds and the mutual information of Alice and Eve will now become independent of $D$, and it may be given as
\begin{align}
\begin{split}
	I^{AE}(|\beta_{\phi_0}|^2,|\beta_{\phi_1}|^2)=& 1+|\beta_{\phi_0}|^2 \log_2 |\beta_{\phi_0}|^2+ |\beta_{\phi_1}|^2 \log_2 |\beta_{\phi_1}|^2
	\label{iae_indep1}
\end{split}
\end{align}
Now, our task is to express $I^{AE}(|\beta_{\phi_0}|^2,|\beta_{\phi_1}|^2)$ in terms of the concurrences $C_{\psi_0}$ and $C_{\chi_0}$. Since $|\beta_{\phi_0}|^2$ and $|\beta_{\phi_1}|^2$ can take two values in terms of concurrences so we need to analyze the relations given in (\ref{cons2}).\\
If possible let $|\beta_{\phi_0}|^2 =\frac{1-\sqrt{1-C_{\chi_0}^{2}}}{2}$. Then (\ref{tau1sq}) and (\ref{cons2}) gives  
\begin{align}
& \left(1-\sqrt{1-C_{\chi_0}^{2}}\right)\left(1-\sqrt{1-C_{\psi_0}^2}\right)=C_{\psi_0}^2\notag \\
&\implies \left(\sqrt{1-C_{\psi_0}^2}-1\right)\left(\sqrt{1-C_{\psi_0}^2}+\sqrt{1-C_{\chi_0}^{2}}\right)=0
\label{equ1}
\end{align}
Since $0<C_{\psi_0}\leq1$ and $0\leq C_{\chi_0}<1$ so (\ref{equ1}) is not possible and we arrive at a contradiction. Thus, $|\beta_{\phi_0}|^2 \neq \frac{1-\sqrt{1-C_{\chi_0}^{2}}}{2}$, Hence, $|\beta_{\phi_0}|^2$ does not belong to $[0,0.5)$.\\
It can be shown that if $|\beta_{\phi_0}|^2 =\frac{1+\sqrt{1-C_{\chi_0}^{2}}}{2}$ then 
\begin{equation}
C_{\psi_0}=C_{\chi_0}
\label{finalequ}
\end{equation}
holds.\\
If $C_{\chi_0}=1$, then $I^{AE}(\frac{1}{2},\frac{1}{2})=0$. So $C_{\chi_0}\neq 1$.
Therefore, alternatively, we can say that the equation (\ref{finalequ}) holds when $|\beta_{\phi_0}|^2\in (0.5,1]$.\\
Similarly, we can also prove that if $|\beta_{\phi_1}|^2\in [0,0.5)$ then
\begin{equation}
C_{\psi_0}=C_{\chi_1}
\label{finalequ1}
\end{equation}
holds.\\
Moreover, one can easily verify that $|\beta_{\phi_1}|^2$ cannot belong to $(0.5,1]$.\\
Therefore, if $|\beta_{\phi_0}|^2\in (0.5,1]$ and $|\beta_{\phi_1}|^2\in [0,0.5)$, then we have 
\begin{equation}
C_{\psi_0}=C_{\chi_0}=C_{\chi_1}
\label{final}
\end{equation}
Hence, using (\ref{iae_indep1}) and (\ref{final}), the expression for $I^{AE}(|\beta_{\phi_0}|^2,|\beta_{\phi_1}|^2)$ can be re-expressed as 
\begin{align}
\begin{split}
	I^{AE}(C_{\psi_0})=& 1+ \frac{1+\sqrt{1-C_{\psi_0}^2}}{2}\log_2 \frac{1+\sqrt{1-C_{\psi_0}^2}}{2} \notag\\
	&+\frac{1-\sqrt{1-C_{\psi_0}^2}}{2} \log_2 \frac{1-\sqrt{1-C_{\psi_0}^2}}{2}
	\label{iae_indep}
\end{split}
\end{align}

\begin{figure}[H]
	\centering
	\includegraphics[width=1\linewidth]{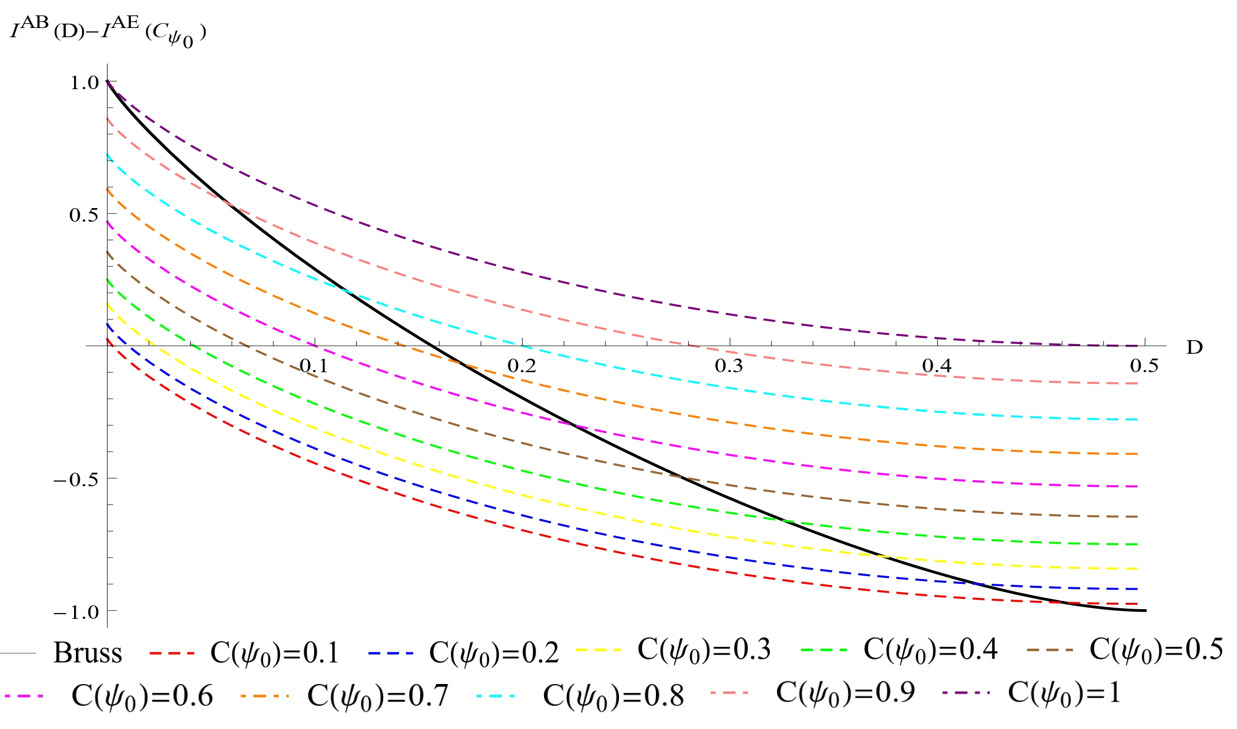}
	\caption{The figure represents that in the modified six-state QKD protocol, the secret key is generated when the curves are lying above $D-$ axis in ($D$,$I^{AB}(D)-I^{AE}(C_{\psi_0})$)-plane and the secret key is not generated when the curves lying below the $D-$ axis. The different curves in the graph are drawn for different values of the concurrence of the ancilla states. The black curve represents when the secret key is generated or is not generated in Bruss's six-state QKD protocol.}
	\label{fig:independent-d-graph}
\end{figure}

The following insights may be drawn from the graph (FIG.\ref{ind d graph}) obtained above:\\
(i) As the value of concurrence $C_{\psi_0}$ increases from 0 to 1, the value of $I^{AB}(D)-I^{AE}(C_{\psi_0})$ increases for increasing value of the disturbance $D$ from 0 to 0.5.\\
(ii) Unlike the model proposed by \cite{bruss_1998}, the secret key may not be generated after a certain value of disturbance $D$, depending upon $C_{\psi_0}\in (0,1]$. However, one may find that the secret key may only be generated in the full range of $D$ when the ancilla state $\ket{\psi_{0}}$ is maximally entangled state.

\section{Conclusion}
To summarize, we have studied a six-state QKD protocol by modifying the unitary transformation prescribed in Bruss's work \cite{bruss_1998}. The modified unitary operation might be able to generate entangled ancilla states at the output, and it might give extra power to the eavesdropper, Eve, to gain knowledge about the key that would be generated between two distant partners, Alice and Bob. The amount of entanglement in ancilla states is measured by the concurrence. We have shown that the mutual information of Alice and Eve not only depends on the disturbance produced while extracting the information but also on the concurrence of the ancilla state and we denote it by $I^{AE}(C_{\chi_0},C_{\chi_1},\tau_1,D)$. Moreover, we have identified the region where the inequality $I^{AB}(D)>I^{AE}(C_{\chi_0},C_{\chi_1},\tau_1,D)$ holds i.e., we have identified the region where the secret key may be generated in spite of the presence of Eve. To analyze this study, we divided the modified protocol into two parts. Firstly, we investigate the case where mutual information of Alice and Eve depends upon both disturbance $D$ and concurrence of the entangled ancilla states. We find that as concurrence increases, the value of $I^{AB}(D)-I^{AE}(C_{\psi_0},C_{\chi_0},D)$ will be positive and increases as the disturbance increases. Further, we noticed in the proposal proposed in \cite{bruss_1998} that the secret key is not generated after a certain value of disturbance $D$, i.e. when $D>0.1565$ , but in our case, the secret key is generated up to a critical value $D_{c}$ of $D$, i.e. the secret key is generated when $0<D_{c}<0.4999999$, otherwise the secret key is not generated, i.e., when $0.4999999 \leq D_{c} <0.5$. In the second case, we find that there exists a relationship between the concurrences of the ancilla states for which the mutual information of Alice and Eve is independent of $D$. In this scenario, the value of $I^{AB}(D)-I^{AE}(C_{\psi_0})$ increases for increasing values of concurrence and the disturbance $D$ respectively. One may also find that in this case, the secret key is not generated by following Bruss's six-state QKD protocol when the disturbance $D>0.1565$ but in our modified six-state QKD protocol, the secret key is generated even for $D>0.1565$ but the concurrence of the ancilla state is adjusted between $(0.74,1)$. The key is generated in the whole domain of $D$ if the ancilla component $\ket{\psi_{0}}$ is a maximally entangled state. To conclude, we feel that it would be interesting to study the same when Alice prepares her state as a mixed state instead of a pure state.\\

\section{Data availability statement}
Data sharing not applicable to this article as no datasets were generated or analysed during the current study.

\section{Author contribution statement}
\textbf{RJ, SA}: Conceptualization, Methodology, Software, Writing- Original draft preparation, Visualization, Writing- Reviewing and Editing.
\balance

\appendix

\section{Derivation of the expression of $\tau_1^2$}

Recalling (\ref{cb}) and squaring both sides, we get
\begin{align}
(1+\tau_1^2)^2 C_{\psi_0}^2=4\tau_1^2
\end{align}
Simplifying and expressing $\tau_1^2$ in terms of the concurrence $C_{\psi_0}$, we get
\begin{align}
\tau_1^2= \frac{2-C_{\psi_0}^2\pm 2\sqrt{1-C_{\psi_0}^2}}{C_{\psi_0}^2}
\end{align}
Now, we will prove that 
\begin{align}
\tau_1^2= \frac{2-C_{\psi_0}^2 - 2\sqrt{1-C_{\psi_0}^2}}{C_{\psi_0}^2}
\end{align}
holds.\\
\textbf{Proof:} If possible, let
\begin{align}
\tau_1^2= \frac{2-C_{\psi_0}^2 + 2\sqrt{1-C_{\psi_0}^2}}{C_{\psi_0}^2}
\end{align}

Since $0< \tau_1^2\leq 1$, so we have
\begin{align}
0<\frac{2-C_{\psi_0}^2+ 2\sqrt{1-C_{\psi_0}^2}}{C_{\psi_0}^2}\leq1
\end{align}
We may observe that LHS inequality is always true but simplifying the RHS inequality, we get
\begin{align}
\sqrt{1-C_{\psi_0}^2}(\sqrt{1-C_{\psi_0}^2}+1)\leq 0
\end{align}
We arrive at a contradiction because $0\leq C_{\psi_0}\leq 1$.\\
Thus, we have 
\begin{align}
\tau_1^2= \frac{2-C_{\psi_0}^2- 2\sqrt{1-C_{\psi_0}^2}}{C_{\psi_0}^2}
\end{align}
Hence proved.

\section{Detailed calculation to derive the specific form of $\ket{\phi_0}$ and $\ket{\phi_1}$}

To start with, let us consider the general form of initially considered pure two-qubit ancilla state $\ket{\phi_0}$ and $\ket{\phi_1}$ in the computational basis $\{\ket{00},\ket{01},\ket{10},\ket{11}\}$ as

\begin{align}
\ket{\phi_0}=&\alpha_{\phi_0} \ket{00}+\beta_{\phi_0} \ket{01}+\gamma_{\phi_0} \ket{10}+\delta_{\phi_0} \ket{11},\notag \\
\text{with}~~&|\alpha_{\phi_0}|^2+|\beta_{\phi_0}|^2+|\gamma_{\phi_0}|^2+|\delta_{\phi_0}|^2=1
\label{a2phi0}
\end{align}
and
\begin{align}
\ket{\phi_1}=&\alpha_{\phi_1} \ket{00}+\beta_{\phi_1} \ket{01}+\gamma_{\phi_1} \ket{10}+\delta_{\phi_1} \ket{11}\notag \\
\text{with}~~&|\alpha_{\phi_1}|^2+|\beta_{\phi_1}|^2+|\gamma_{\phi_1}|^2+|\delta_{\phi_1}|^2=1
\label{a2phi1}
\end{align}
Further, we choose the states $\ket{\psi_0}$ and $\ket{\psi_1}$ in such a way that ($\ref{bd}$), i.e., $\braket{\psi_0}{\psi_1}=0$ is satisfied. Therefore, we can consider the following form of the states
\begin{align}
\ket{\psi_0}&=\frac{\ket{00}-\tau_1\ket{11}}{\sqrt{1+\tau_1^2}}
\label{a2psi0}
\end{align}
and 
\begin{align}
\ket{\psi_1}&=\frac{\tau_1\ket{00}+\ket{11}}{\sqrt{1+\tau_1^2}}
\label{a2psi1}
\end{align}
where $0< \tau_1 \leq 1$.\\
Let us now calculate the disturbance made by Eve if Alice prepares the state in the basis $B_{1}$: 
\begin{align}
\bra{0}\rho_B\ket{0}&=1-D\\
\bra{1}\rho_B\ket{1}&=1-D
\end{align}
where, $\rho_B=\text{Tr}_{AE_1E_2} (\rho_{ABE_1E_2})$ represents the state sent by Alice to Bob which is intercepted by Eve. \\
In the same way, if Alice prepares the state in the bases $B_{2}$ or $B_{3}$ and keeping in mind the restrictions that make the disturbance same for all the states in the three non-orthogonal bases $\{B_{1},B_{2},B_{3}\}$, then for $0<\tau_1\leq 1$, we have  
\begin{align}
(\alpha_{\phi_0}+\alpha^*_{\phi_1})(1-\tau_1)+  (\delta_{\phi_0}+\delta^*_{\phi_1})(1+\tau_1)&=0\label{0pm1byr2}\\
(\alpha_{\phi_0}+\alpha^*_{\phi_1})(1+\tau_1)+  (\delta_{\phi_0}+\delta^*_{\phi_1})(1-\tau_1)&=0 \label{0pmi1byr2}
\end{align}
Solving (\ref{0pm1byr2}-\ref{0pmi1byr2}), we obtain 
\begin{align}
\delta_{\phi_0}+\delta^*_{\phi_1}=0 \label{a2cond1}\\
\alpha_{\phi_0}+\alpha^*_{\phi_1}=0 \label{a2cond2}
\end{align}
Recalling (\ref{unitaryU}) and (\ref{constraint}), and using (\ref{a2phi0}-\ref{a2psi1}), we get
\begin{equation}
\begin{split}
	\tau_1(\alpha^*_{\phi_0}-\delta_{\phi_1})+\delta^*_{\phi_0}+\alpha_{\phi_1}&=0\\
	\tau_1(\alpha_{\phi_1}-\delta^*_{\phi_0})+\delta_{\phi_1}+\alpha_{\phi^*_0}&=0
	\label{a2tau1}
\end{split}
\end{equation}
To make (\ref{a2tau1}) independent of $\tau_1$, we have
\begin{align}
\alpha^*_{\phi_0}-\delta_{\phi_1}=0\label{a2cond3}\\
\alpha_{\phi_1}-\delta^*_{\phi_0}=0\label{a2cond4}
\end{align}
Using (\ref{a2cond3}-\ref{a2cond4}), equation (\ref{a2tau1}) reduces to
\begin{align}
\delta^*_{\phi_0}+\alpha_{\phi_1}=0\label{a2cond5}\\
\delta_{\phi_1}+\alpha^*_{\phi_0}=0\label{a2cond6}
\end{align}
Solving (\ref{a2cond3}) \& (\ref{a2cond6}) and (\ref{a2cond4}) \& (\ref{a2cond5}), we get
\begin{align}
\alpha_{\phi_0}=\delta_{\phi_0}=\alpha_{\phi_1}=\delta_{\phi_1}=0
\label{b16}
\end{align}
Using (\ref{b16}), equation (\ref{a2phi0}) and (\ref{a2phi1}) reduces to 
\begin{align*}
\ket{\chi_0}&= \beta_{\phi_0} \ket{01}+\gamma_{\phi_0}\ket{10}, ~ |\beta_{\phi_0}|^2+|\gamma_{\phi_0}|^2=1\\
\ket{\chi_1}&= \beta_{\phi_1} \ket{01}+\gamma_{\phi_1}\ket{10}, ~ |\beta_{\phi_1}|^2+|\gamma_{\phi_1}|^2=1 
\end{align*}

\end{document}